\documentclass[aps,prb,amsmath,amssymb,twocolumn,superscriptaddress,showpacs,floatfix]{revtex4-1}

\DeclareMathOperator{\TCFE}{\mathit{T}_{\mathrm{\scriptscriptstyle C}}^{\mathrm{\scriptscriptstyle FE}}}
\DeclareMathOperator{\taup}{\tau_{\mathrm{\scriptscriptstyle P}}}

\DeclareMathOperator{\Eex }{\mathit E_{\mathrm{ext}}}

\DeclareMathOperator{\epsFE }{{\epsilon}^{\mathrm{\scriptscriptstyle FE}}}

\usepackage{graphicx}% Include figure files
\usepackage{dcolumn}% Align table columns on decimal point
\usepackage{bm}% bold math

\usepackage{amsmath}
\usepackage{amssymb}

\begin{document}

% Use the \preprint command to place your local institutional report
% number in the upper righthand corner of the title page in preprint mode.
% Multiple \preprint commands are allowed.
% Use the 'preprintnumbers' class option to override journal defaults
% to display numbers if necessary
%\preprint{}

%Title of paper
\title{Proximity coupling of granular film with ferroelectric substrate and giant electro-resistance effect}

\author{O.~G.~Udalov}
\affiliation{Department of Physics and Astronomy, California State University Northridge, Northridge, CA 91330, USA}
\affiliation{Institute for Physics of Microstructures, Russian Academy of Science, Nizhny Novgorod, 603950, Russia}

\author{N.~M.~Chtchelkatchev}
\affiliation{Department of Physics and Astronomy, California State University Northridge, Northridge, CA 91330, USA}
\affiliation{L.D. Landau Institute for Theoretical Physics, Russian Academy of Sciences,117940 Moscow, Russia}
\affiliation{Department of Theoretical Physics, Moscow Institute of Physics and Technology, 141700 Moscow, Russia}

\author{I.~S.~Beloborodov}
\affiliation{Department of Physics and Astronomy, California State University Northridge, Northridge, CA 91330, USA}

\date{\today}


\begin{abstract}
We study electron transport in granular film placed above the ferroelectric substrate.
We show that the conductivity of granular film strongly depends on the ferroelectric state
due to screening effects which modify the Coulomb blockade in granular film.
In particular, the electric current in granular film is controlled by the direction of
ferroelectric polarization. We show that the ferroelectric/granular film system has a large
electro-resistance effect. This effect can be utilized
in memory and electric field sensor applications.
\end{abstract}

%\maketitle must follow title, authors, abstract, \pacs, and \keywords
\maketitle

\section{Introduction\label{sec:intro}}

Ferroelectric materials were known for a long time, nearly a century. Currently ferroelectric materials
attract a lot of attention for two reasons: on one hand, due to the availability
of new kind of ferroelectrics (FE) for experiment including ultrathin films, lateral confined and hybrid ferroelectric nanoparticles,
and granular ferroelectrics;~\cite{Zlatkin1998,Fu2004,Frid2010rev,Scott2007,Jonas2009} on the other hand, due to the
promising potential applications of new ferroelectric materials. New physics emerges in these materials compared to
the bulk ferroelectric materials investigated in the past.
Electric properties of ultrathin films are qualitatively different from their bulk counterpartners.~\cite{Frid2006rev,Frid2010rev} New phases such as electric vortices appear in ferroelectric nanoparticles.~\cite{Zlatkin1998} Combination of ferroelectric films with ferromagnetic layers produces a strong strain and charge mediated magneto-electric coupling.~\cite{Scott2006,Barthelemy2010} All these
discoveries are giving a new lease of life to this field.

The fundamental question in this field is the influence of the FE polarization on transport properties.
This issue is crucial for memory applications since the direction of the FE polarization can be used
for data coding. In this case the mechanism of data
writing involves the electric field rather than the electric current.
This is the advantage of the FE based memory allowing to essentially reduce the energy consumption.
\begin{figure}[tb]
\includegraphics[width=0.7\columnwidth]{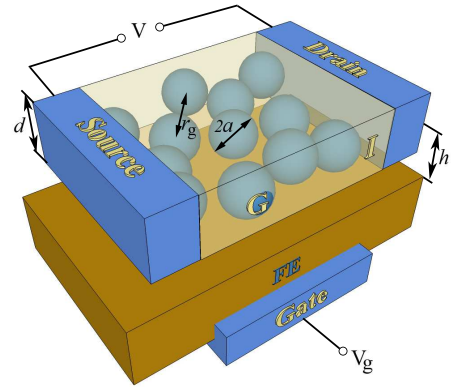}
\caption{(Color online) Sketch of ferroelectric/granular film system with
granular film of thickness $d$ being placed at distance $h$ above the ferroelectric (FE) layer. Granular film
consists of metallic grains (G) with average size $2a$ and intergrain distance  $r_{\textrm{g}}$
embedded into an insulating matrix (I). $V$ and $V_\textrm{g}$ are the bias and the gate
voltages, respectively.}\label{Fig_sys}
\end{figure}

There are several ways to determine the direction of the FE polarization:
1) the tunneling electro-resistance effect in FE
tunnel junctions with tunneling current being dependent on the
direction of the FE polarization.~\cite{Kohlstedt2006,Alexe2010,Tsy2005}
This method requires very thin FE barriers, thus reducing the FE Curie temperature
and the electric polarization of the FE barrier. 2) The rectification effect in the Shottky diode at the
ferroelectric/metal interface.~\cite{Krijn1994,Tsymbal2013,Cheong2009,Kalinin2009}
This effect changes the sign by switching the FE polarization. 3) The transistor type structures with
FE being placed in between the gate electrode and the transistor channel. This method uses electric charges
on the surface of the FE to control the current in the channel.~\cite{HLILMEIER1966, McWhorter1992,Benedetto1997}
Unfortunately the
complexity of "good" FE/semiconductor surfaces does not allow the realization of this type of memory now.~\cite{Benedetto1997}

In this paper we introduce a new mechanism for finding the direction of the FE polarization
using the granular thin film, with metallic grains embedded in an insulating matrix, for transistor channel, Fig.~\ref{Fig_sys}.
We assume that grains have nanometer range sizes with well developed Coulomb blockade
in granular film. The main parameter characterizing the Coulomb blockade
in granular materials is the charging energy which is the electrostatic energy
of a single excess electron placed on a single grain. Charging the charging energy one can
control the electron transport in granular systems.

Recently it was shown that the FE matrix influences the charging energy
in granular ferroelectrics -- materials consisting of metallic grains embedded
into a FE matrix.~\cite{Bel2014} In this paper we propose an
essentially different geometry: the FE film being placed between the gate electrode
and the channel/granular film, Fig.~\ref{Fig_sys}. Since the Coulomb interaction is
the long range interaction the presence of the FE layer will influence the charging energy
in granular film and will define the electron transport in granular channel. In contrast to the
field effect transistor (FET) with semiconducting channel, where the FE polarization changes the position of the chemical potential in the channel leaving the band gap unchanged, in the FET with granular channel the ferroelectric dielectric permittivity influences the width of the "forbidden band" (determined by the charging energy).

The chosen geometry in Fig.~\ref{Fig_sys} allows:
1) to control the FE state using the gate voltage without inducing the electric current in the system; 2) to identify the FE state using the current along the granular film without affecting the FE substrate;
and 3) to avoid surface problems existing in the case of semiconductor/FE transistor.

In this paper we show that: 1) the source to drain current in the system in Fig.~\ref{Fig_sys} depends on the direction of
the FE polarization and the gate voltage and 2) the temperature and electric field dependence of dielectric permittivity
can be studied using the conductivity of granular film.

The paper is organized as follows: In Sec.~\ref{sec:sys}-\ref{sec:sys_2} we introduce the model and discuss the influence of dielectric substrate on the Coulomb energy of granular film.
We study transport properties of composite system in Fig.~\ref{Fig_sys} in Sec.~\ref{sec_transp1}-\ref{sec_transp3}
and discuss the validity of our model and some possible applications in the Discussion section.

\section{Influence of ferroelectricity on transport properties of granular film \label{sec:Tran}}

\subsection{The system\label{sec:sys}}

We consider the following system: the quasi two dimensional granular film of thickness $d$ placed above
the FE substrate at distance $h$ ($d < h$, $h$ is the distance between the grain centres and the
FE surface), Fig.~\ref{Fig_sys}.  The film consists of metallic grains with average radius $a$ embedded
in an insulating matrix ($d\approx 2a$). The average intergrain distance is $r_{\textrm g}$. The film is confined with two leads - source and drain.
A small voltage bias $V$ is applied to the leads leading to the electric current $\mathrm{j}$.
There is an insulating layer with dielectric constant $\epsilon_{\textrm I}$ located in between the granular
film and the FE substrate. The FE substrate has the thickness $L\gg h$.
The ferroelectric state is controlled by the gate electrode with voltage $V_{\mathrm{g}}$ placed beneath the FE substrate.
This gate creates an external electric field $\Eex$ inside the FE layer perpendicular to the film and the substrate.
This external field $\Eex$ does not create the electric current.

\subsection{Ground state\label{sec:sys_1}}

For zero bias voltage the electric current is absent in the system. The FE
substrate is affected by the electric field created by the gate electrode, grains, and charges
existing on the FE surfaces. We assume that the granular film has small but finite conductivity.
Therefore the electric field created by the surface charges of the FE is screened by the grains charges.
Also we assume that the measurement time is much larger than any relaxation times in the system.
Thus, all grains have the same electric potential and the field inside the FE is screened due
to presence of grains~\cite{Shklowski2014}.
For distances $L\epsilon_{\mathrm I}\gg (h-a)r^2_{\mathrm g} \epsFE/(ha)$ the screening
charge at a grain at zero gate voltage is $q^{\mathrm g}=-Pr_{\mathrm g}^2$ with $P$ being the magnitude of
electric polarization, Appendix~\ref{sec:AppScren}.
$\epsFE=(\epsilon^{\mathrm{FE}}_{\perp}\epsilon^{\mathrm{FE}}_{||})^{1/2}$ is the effective dielectric permittivity of the FE substrate with $\epsilon^{\mathrm{FE}}_{\perp,||}$ being the dielectric permittivity tensor components perpendicular and parallel to the
FE surface.\cite{Mele2001} It depends on the electric field $\Eex$ applied to the FE.
Corrections to the charge $q^{\mathrm g}$ are of the order
of $(h-a)r^2_{\mathrm g}\epsFE/(L\epsilon_{\mathrm I}ha)$. In this limit the
external field can be written in the form, $\Eex=V_{\mathrm g}/L$.

At finite gate voltage with similar accuracy we find $q^{\mathrm g}=(-P+V_{\mathrm g}\epsFE/(4\pi L))r_{\mathrm g}^2$.

We notice that the charge $q^{\mathrm g}$ is the average grain charge. In real systems due to electron charge quantization
$q^{\mathrm g}$ can have integer values only for each individual grain. This leads to an
additional source of disorder in the system. The polarization of typical
FE's gives $q^{\mathrm g}\gg 1$ $e$ for grain sizes of 5 nm. Therefore we neglect this charge quantization here.

\subsection{Transport model\label{sec:sys_2}}

The FET with semiconductor channel and with the
FE in between the gate and the channel is well studied.~\cite{HLILMEIER1966, McWhorter1992,Benedetto1997,Desu2004,Chin2004}
In this device the FE layer controls the electron density in the channel.
Screening of the FE polarization by the charges accumulated in the channel
leads either to enrichment or depletion of the channel and correspondingly either to
increasing or decreasing of the current in the channel.

In the current consideration the granular film plays the role of the channel. Screening of the
FE polarization leads to charging of metallic grains inside the channel. Since metallic
grains have large number of electrons an addition of several electrons does not
influence the electron transport in contrast to the semiconductor case.~\cite{Shklowski2014}

The most important parameter characterizing the properties of
granular film is the charging energy $E_{\mathrm{c}}$,~\cite{Arie1973,Bel2007review}.
In granular metals it plays a similar role as the forbidden band in semiconductors.
For charging energy $E_{\mathrm{c}}\gg T$ electrons are localized inside grains.
The Coulomb energy in granular materials depends on the state of the
FE due to screening effects.~\cite{Bel2014,Beloborodov2014,Beloborodov2014_1}
The perpendicular component of the dielectric permittivity $\epsFE$ of the FE can be
changed using the external electric  field and temperature leading to the controllable charging energy.
For example, the Coulomb energy of granular metal placed above the FE substrate has the form,
Appendix~\ref{Sec:App_Coulomb}~\cite{Beloborodov2014}
\begin{equation}\label{Gap}
E_{\mathrm{c}}=E_{\mathrm{c}}^0\frac{1}{\epsilon}\left(1+\frac{a}{2h}\frac{\epsilon-\epsFE}{\epsilon+\epsFE}\right),
\end{equation}
where $E_{\mathrm{c}}^0=e^2/2a$ is the Coulomb energy of the charged sphere in vacuum and
$\epsilon$ is the effective dielectric permittivity of the half space above the FE substrate. The
estimates of $\epsilon$ is given in Appendix~\ref{Sec:App_Coulomb}.  In our consideration
$\epsilon$ does not depend on the external electric field.
It was shown in Ref.~\onlinecite{Beloborodov2014,Beloborodov2014} that the FE layer is strongly influence
the magnetic state of the proximity coupled granular film with
magnetic grains through screening effects. Here we uncover the new physics related to the electron transport in the
proximity coupled granular nonmagnetic film and the FE layer shown in Fig.~\ref{Fig_sys}.

To summarize, in contrast to the FET with semiconducting channel,
where screening of the FE polarization changes the chemical potential in the channel leaving the band gap unchanged,
in granular metals the changing of the FE dielectric permittivity modifies the forbidden band width in the channel.
\begin{figure}[t]
\includegraphics[width=1\columnwidth]{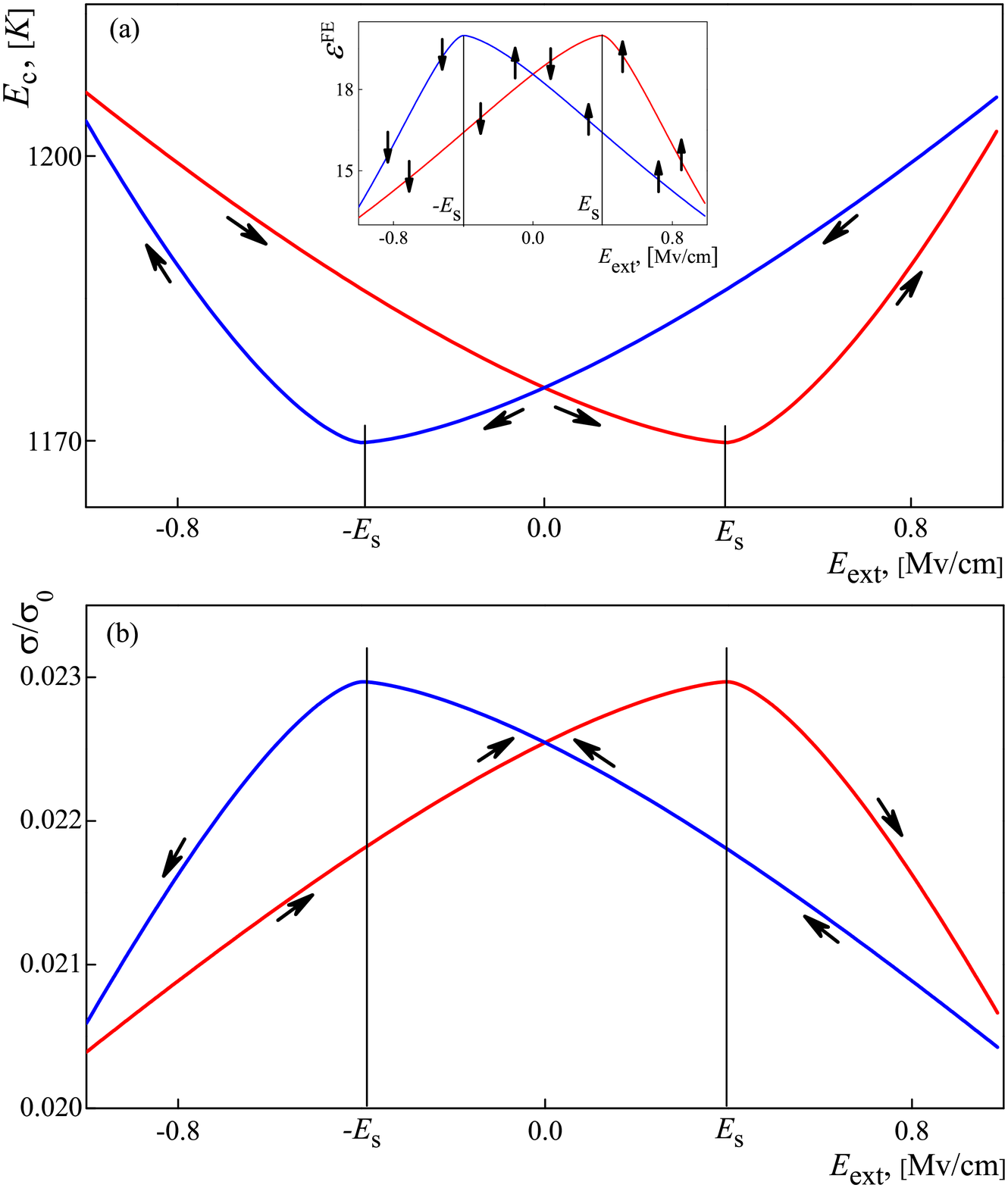}\\
\caption{(Color online) (a) Charging energy $E_c$
in Eq.~(\ref{Gap}) vs external electric field $\Eex$ for ferroelectric/granular film
system shown in Fig.~\ref{Fig_sys}. $E_{\mathrm{s}}$ is the ferroelectric switching field.  Arrows indicate the path around the hysteresis loop. Insert: dielectric permittivity vs field $\Eex$ for temperatures $T < \TCFE$. Vertical arrows
indicate the direction of polarization of the substrate. (b) Normalized conductivity of granular film vs external field $\Eex$.
Arrows indicate the path around the hysteresis loop. The following set of parameters were used: the average grain size $a=3$~nm, the intergrain distance $r_{\textrm g}=7$ nm, the dielectric permittivity of insulating matrix $\epsilon_{\mathrm{m}}=1$, the effective dielectric permittivity of the half-space above the substrate $\epsilon=1.3$, the distance between the film and the substrate $h=4$~nm.}\label{Cond}
\end{figure}

\subsubsection{Variable range hopping}

There are several transport regimes in composite materials depending on the coupling between the grains.
At weak coupling and low temperatures the electron transport is due to variable range hopping.
This mechanism involves electrons with energies inside the Coulomb gap leading to the following
conductivity~\cite{Shklovskii1975,Shklovskii2012,Vinokur2005}
\begin{equation} \label{Eq_Cond_Cot}
\sigma = \sigma_0 \exp(-(T_0/T)^{1/2}).
\end{equation}
Here $\sigma_0$ is the high temperature conductivity proportional to the intergrain conductance.
In our consideration the system is two dimensional since
the film thickness is comparable with a single grain diameter and the hopping distance at low
temperatures is larger than the grain size. $T_0$ is the characteristic temperature scale
\begin{eqnarray} \label{Eq_Temp}
T_0 = e^2/(\epsilon\xi),
\end{eqnarray}
where $\xi$ is the electron localization length~\cite{Bel2007review}
\begin{equation} \label{Eq_Loc}
\xi=a/\ln(E^{2}_{c}/T^2g^{0}_{t}).
\end{equation}
$g^0_t$ is the intergrain tunneling conductance.

\subsubsection{Sequential tunneling}

Increasing the temperature the hopping distance decreases reaching the grain size.
At these temperatures and small bias voltage the conductivity has an activation
behavior in the Coulomb blockade regime~\cite{Bel2007review}
\begin{equation}\label{Conduct}
\sigma=\sigma_0e^{-\frac{E_{\mathrm{c}}}{T}},
\end{equation}
with $\sigma_0$ being the high temperature conductivity proportional to the intergrain conductance.

Using Eqs.~(\ref{Eq_Cond_Cot}) and (\ref{Conduct}) we find that for temperatures $T>T_{\mathrm{cr}}=E_{\mathrm c}^2/T_0$
the activation transport is more important. Neglecting the logarithmic dependence of the localization length 
$\xi$ on temperature for the following set of parameters 
$E_\mathrm c\sim10^3$ K, $g^0_{t}=0.1$, $E_\mathrm c^0\sim 1.7\cdot 10^3$ K
we find $T_{\mathrm{cr}}\approx 100$ K. Therefore at room temperature, $T > T_{\mathrm{cr}}$ the 
sequential tunneling is the main transport mechanism. For FE with room temperature transition, the 
transport is described by Eq.~\ref{Conduct}. For FE with Curie temperature less than $T_{\mathrm{cr}}$ 
one has to use Eq.~\ref{Eq_Cond_Cot}.

Below we mostly discuss the high temperature limit since the FE Curie temperature is
rather high. However, we also provide the necessary formulas for low temperatures.

The Coulomb energy $E_{\mathrm{c}}$ in Eq.~(\ref{Conduct}) depends on the dielectric permittivity  $\epsFE$ of the FE substrate which in its turn depends on the external electric field $\Eex$ and temperature. Thus, applying the gate voltage $V_{\textrm{g}}$ and therefore the external field $\Eex$ we can control the conductivity of granular film. Next, we discuss two conductivity regimes --- below and above
the ferroelectric Curie temperature $\TCFE$.
\begin{figure}[t]
\includegraphics[width=1\columnwidth]{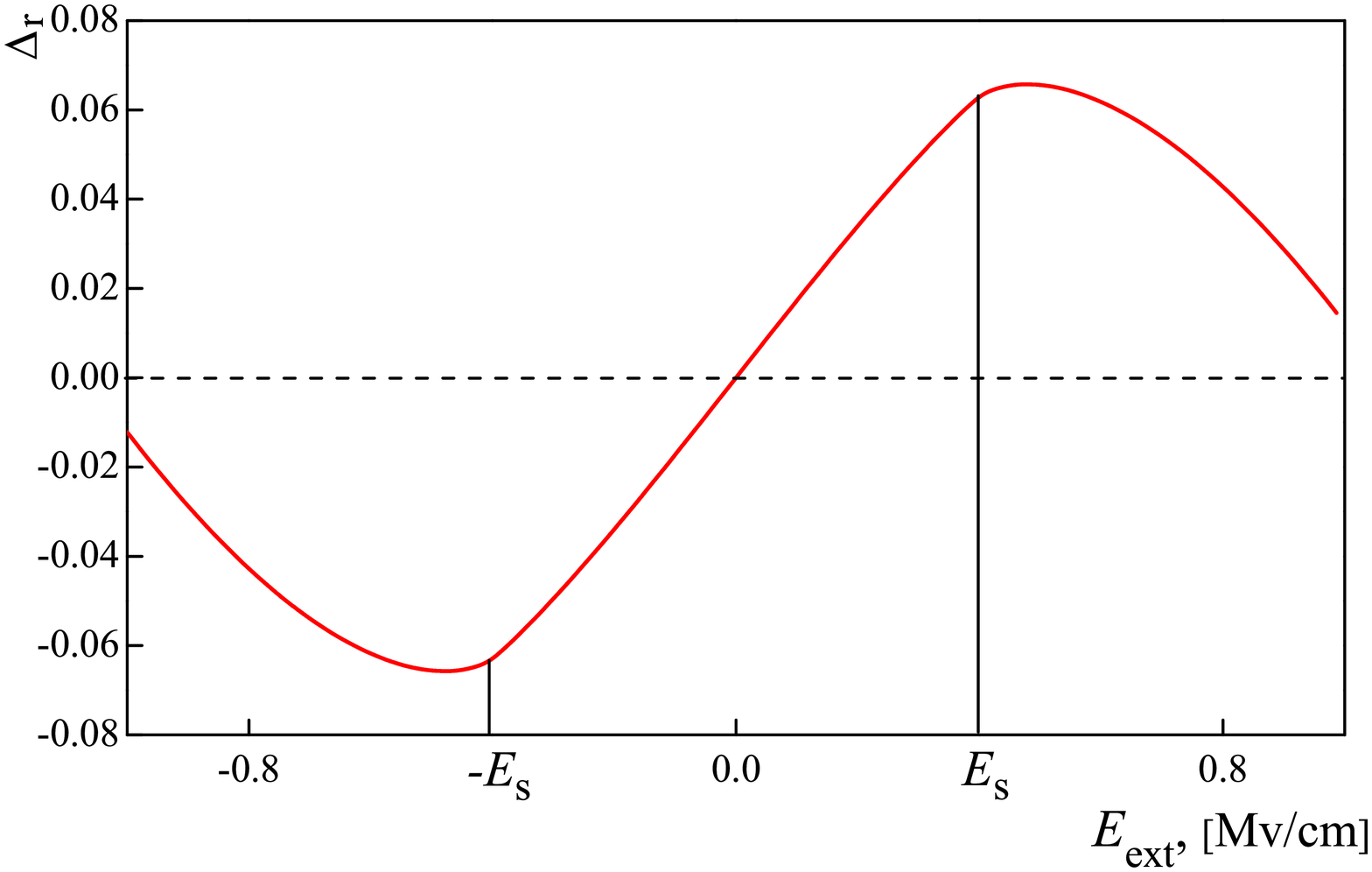}\\
\caption{(Color online) Dimensionless conductivity difference  $\Delta_{\mathrm{r}}$ in Eq.~(\ref{Eq_ConDif}) of granular film placed above the FE substrate vs external electric field. $E_{\mathrm{s}}$ is the FE switching field. }\label{ConDif}
\end{figure}

\subsection{Conductivity below the Curie point \label{sec_transp1}}

Here we discuss the conductivity of granular film for temperatures $T < \TCFE$ with
the FE substrate being in the FE state. In this case the polarization and the dielectric permittivity
$\epsFE$ of the FE layer have hysteresis behavior as a function of external field $\Eex$ meaning that
there are two different resistive states for each field depending on the hysteresis branch.

The Coulomb energy  $E_{\mathrm{c}}$ in Eq.~(\ref{Gap}) depends on
the dielectric permittivity $\epsilon^{\rm\scriptscriptstyle{FE}}$
and therefore shows the hysteresis behavior, Fig.~\ref{Cond}(a).  It has a minimum for external
fields close to the switching field $E_{\mathrm{s}}$. The behavior of dielectric permittivity vs external field at fixed temperature $T=310$ K is shown in the inset in Fig.~\ref{Cond}(a). We assume that the
granular film is placed above the P(VDF/TrFE) relaxor FE. For the dependence of dielectric permittivity of
P(VDF/TrFE) on the external field we use the data of Ref.~\onlinecite{Park2006} with some smooth function
to describe the data. The granular film has the following parameters:
the average grain size $a=3$~nm, the intergrain distance $r_{\textrm g}=7$ nm, the dielectric permittivity of insulating matrix $\epsilon_{\mathrm{m}}=1$, the effective dielectric permittivity of the half-space above the substrate $\epsilon=1.3$, the distance between the film and the substrate $h=4$~nm.

The hysteresis in the charging energy leads to the hysteresis in the resistance of granular film, Fig.~\ref{Cond}(b). The resistance difference between the hysteresis branches is important for memory applications. The maximum difference occurs in the vicinity of the switching field $E_{\mathrm{s}}$. We introduce the dielectric permittivity of the
FE substrate for these two states as $\epsilon^{\rm\scriptscriptstyle{FE}}_1$ and $\epsilon^{\rm\scriptscriptstyle{FE}}_2$. Substituting Eq.~(\ref{Gap}) into Eq.~(\ref{Conduct}) we find
\begin{equation}\label{Eq_ConDif}
\Delta_{\mathrm{r}}=2\frac{\sigma(\epsilon^{\rm\scriptscriptstyle{FE}}_1)-\sigma(\epsilon^{\rm\scriptscriptstyle{FE}}_2)}{\sigma(\epsilon^{\rm\scriptscriptstyle{FE}}_1)+
\sigma(\epsilon^{\rm\scriptscriptstyle{FE}}_2)}=
\tanh\left(\frac{2e^2}{hT}\frac{\epsilon^{\rm\scriptscriptstyle{FE}}_1-\epsilon^{\rm\scriptscriptstyle{FE}}_2}
{(\epsilon^{\rm\scriptscriptstyle{FE}}_1+\epsilon)(\epsilon^{\rm\scriptscriptstyle{FE}}_2+\epsilon)}\right).
\end{equation}
\begin{figure}[t]
\includegraphics[width=1\columnwidth]{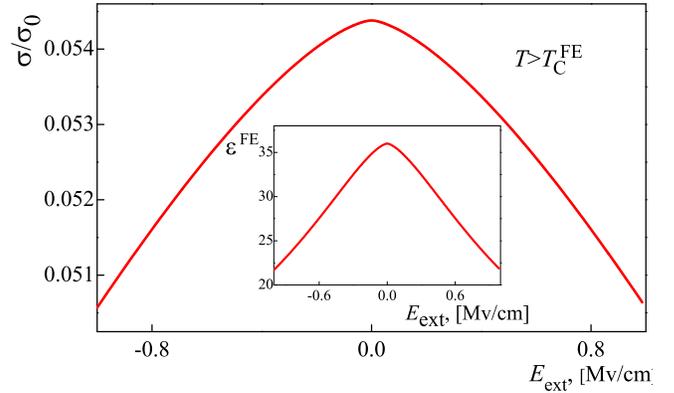}\\
\caption{(Color online) Normalized conductivity of granular film vs external
electric field $\Eex$ for temperatures $T = 390K > \TCFE$. Insert: Dielectric permittivity
of the FE substrate vs external electric field.}\label{ConPE}
\end{figure}

The conductivity difference in Eq.~(\ref{Eq_ConDif}) increases with decreasing the film thickness. The minimum thickness for considered system is the grain size, $h=a$. The conductivity difference depends on the external electric field $\Eex$
and can be rather big, Fig.~\ref{ConDif}.

The conductivities ratio for different FE dielectric permittivity can be estimated as  $\sigma(\epsilon^{\rm\scriptscriptstyle{FE}}_2)/\sigma(\epsilon^{\rm\scriptscriptstyle{FE}}_1)\approx \exp(-(e^2/Th)((\epsilon^{\rm\scriptscriptstyle{FE}}_2-\epsilon^{\rm\scriptscriptstyle{FE}}_1)/ (\epsilon^{\rm\scriptscriptstyle{FE}}_2\epsilon^{\rm\scriptscriptstyle{FE}}_1)))$. At room temperature the factor
$e^2/Th$ is $\sim 5-10$. The second factor in the exponent is less than $1$ and can reach $0.3$.
Therefore the maximum ratio $\sigma(\epsilon^{\rm\scriptscriptstyle{FE}}_2)/\sigma(\epsilon^{\rm\scriptscriptstyle{FE}}_1)$
is about $10$. In this paper this ratio is less due to the small difference between $\epsilon^{\rm\scriptscriptstyle{FE}}_2$ and $\epsilon^{\rm\scriptscriptstyle{FE}}_1$.

In the low temperature limit, $T < T_{\mathrm{cr}}$, where the VRH is the main transport mechanism 
the conductivity difference should be calculated using Eq.~(\ref{Eq_Cond_Cot}). 
In this case $\Delta_{\mathrm r}$ can not be written in a simple form. 
Generally, the conductivity difference is given by the expression
\begin{equation}\label{Eq_ConDif_VHR}
\Delta_\mathrm r=\tanh\left(\frac{\sqrt{T_0(\epsilon^{\rm\scriptscriptstyle{FE}}_2)}-\sqrt{T_0(\epsilon^{\rm\scriptscriptstyle{FE}}_1)}}{2\sqrt{T}}\right).
\end{equation}

\subsection{Conductivity above the Curie point \label{sec_transp2}}

For temperatures $T > \TCFE$ the spontaneous polarization in the FE layer is absent. In this limit the dielectric permittivity is unambiguous function of the external electric field $\Eex$ with no hysteresis. It monotonically decreases with electric field. The conductivity $\sigma$ has similar behavior, Fig.~\ref{ConPE}. Plots
in Fig.~\ref{ConPE} correspond to the granular film placed above the P(VDF/TrFE) relaxor FE substrate
with the same parameters as discussed before.

\subsection{Conductivity temperature dependence \label{sec_transp3}}

\begin{figure}[t]
\includegraphics[width=1\columnwidth]{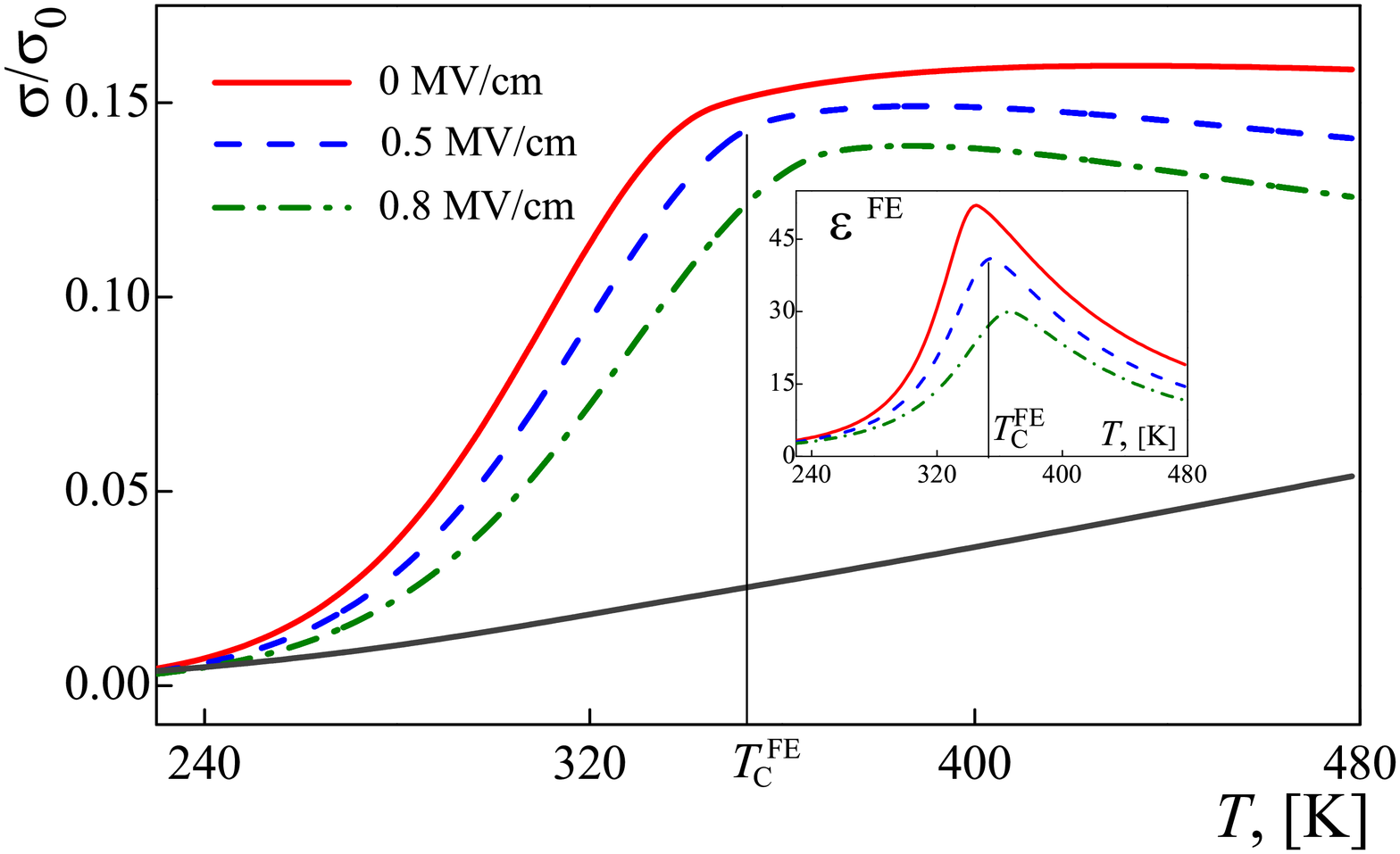}\\
\caption{(Color online) Normalized conductivity of granular film placed above the P(VDF/TrFE) substrate vs temperature
for different electric fields. Insert: dielectric permittivity of the FE substrate vs temperature.
The black solid line stands for conductivity of granular film without FE substrate.}\label{Fig_ConT}
\end{figure}

The temperature dependence of conductivity of granular film placed above the FE substrate is shown in Fig.~\ref{Fig_ConT}.
Different curves correspond to different external electric fields $\Eex$. Three upper curves correspond to granular film placed
above the substrate at distance $h = 4$ nm and
with average grain size $4$ nm. The lower solid line stands for conductivity of granular film without FE substrate. We use the data of Ref.~\onlinecite{Park2006} for dielectric
permittivity of the FE substrate.  The temperature dependence of dielectric permittivity is essentially different from the Weiss law since P(VDF/TrFE) is the relaxor ferroelectric.~\cite{Smolensky1970,Guo1996,Bhalla1998}

The dielectric permittivity $\epsFE$ of the FE substrate strongly depends on temperature in the vicinity of the
FE-paraelectric phase transition, insert in Fig.~\ref{Fig_ConT}. It has a maximum at temperature $T = \TCFE$ leading to the maximum in conductivity. The maximum appears due to the suppression of Coulomb blockade. This is in contrast to the usual situation where conductivity decreases in the vicinity of phase transition due to the increase of current carriers scattering by the order parameter fluctuations.

The change of conductivity on temperature has the form
\begin{equation}\label{Eq_ConSenT}
\frac{\partial \sigma}{\partial T}=\frac{\sigma E^0_{\textrm{c}}}{T}\left\{\frac{1}{\epsilon T}\left(1+\frac{a(\epsilon-\epsFE)}{h(\epsilon+\epsFE)}\right)+\frac{2a(\epsFE)^{\prime}_T}{h(\epsilon+\epsFE)^2}\right\}.
\end{equation}
For large dielectric permittivity of the FE substrate, $\epsFE\gg\epsilon$, we find
\begin{equation}\label{Eq_ConSenTLim}
\frac{\partial \sigma}{\partial T}=\frac{\sigma E^0_{\textrm{c}}}{T}\left\{\frac{1}{\epsilon T}\left(1-\frac{a}{h}\right)+\frac{2a(\epsFE)^{\prime}_T}{h(\epsFE)^2}\right\}.
\end{equation}

Using phenomenological theory of phase transitions we write the dielectric permittivity of conventional ferroelectrics
%(not relaxors, which we discussed in the previous sections)
as follows,~\cite{Levan1983}
\begin{equation}\label{Eq_DielPerm}
\epsFE\approx\left\{\begin{array}{l} \frac{1}{\alpha (T-\TCFE)}, \hspace{0.9cm}T>\TCFE, \\ \frac{1}{2\alpha (\TCFE-T)}, \hspace{0.9cm} T<\TCFE.\end{array}\right.
\end{equation}
Here $\alpha$ is the phenomenological constant.~\cite{Levan1983} Substituting Eq.~(\ref{Eq_DielPerm}) into Eq.~(\ref{Eq_ConSenT}) we obtain
\begin{equation}\label{Eq_ConSenTLim1}
\frac{\partial \sigma}{\partial T}=\frac{\sigma E^0_{\textrm{c}}}{T}\left\{\begin{array}{l}\frac{1}{\epsilon T}\left(1-\frac{a}{h}\right)-\frac{2a\alpha/h}{\left(T-(\TCFE)^2\right)\alpha^2\epsilon+1}, \hspace{0.1cm}T>\TCFE, \\ \frac{1}{\epsilon T}\left(1-\frac{a}{h}\right)+\frac{4a\alpha/h}{\left(T-(\TCFE)^2\right)4\alpha^2\epsilon+1},\hspace{0.1cm}T<\TCFE.\end{array}\right.
\end{equation}
The temperature derivative of conductivity is discontinuous at the transition point leading to the kink in conductivity. In reality the kink is smeared because the phenomenological
Landau-Ginzburg-Devonshire theory is valid only beyond the Ginzburg region.~\cite{Landau5}

The influence of the FE substrate is more pronounced for temperatures $T < \TCFE$, below the FE transition,
if the following condition is satisfied
\begin{equation}\label{Crit1}
\frac{2a}{h}\epsilon\alpha T>1.
\end{equation}
Equation~(\ref{Crit1}) has transparent physical meaning: the dielectric permittivity of
FE should be smaller than the dielectric permittivity of granular
film $\epsilon$ at distance $1/(\alpha T)$ away from the phase transition.

In the low temperature limit, $T<T_{\mathrm{cr}}$, the change of conductivity on temperature is defined by the expression
\begin{equation}\label{Eq_Cond_tem_VRH}
\frac{\partial \sigma}{\partial T}=\frac{\sigma}{\sqrt{TT_0}}\left(\frac{2e^2}{\epsilon a}\left(\frac{2E_{\mathrm c}^0 a\epsilon(\epsFE)^{\prime}_T}{E_\mathrm c h(\epsilon+\epsFE)^2}+\frac{1}{T}\right)+\frac{T_0}{T}\right).
\end{equation}

Recently the conductivity of Pt based nanogranular metal placed above the TTF-CA
organic ferroelectric was studied experimentally.~\cite{Keller2014} The data show a peculiar behavior
of granular film conductivity on temperature in the vicinity of the FE Curie point. The
conductivity has a maximum at temperature $T = \TCFE$. These results are in a good
agreement with our predictions: the conductivity peak exists due to the decreasing of
charging energy in granular film in the vicinity of the Curie temperature, $\TCFE$.

\section{Discussion\label{Sec:App}}

\subsection{Memory cell}

Composite system of granular and FE films can be used as a memory cell with data storage
in the direction of polarization of the FE film.

Writing of data (switching between different states) is
achieved by the external electric field, $\Eex$ or the gate voltage, Fig.~\ref{Fig_sys}. The field $\Eex$
should be larger than the FE switching field $E_{\mathrm{s}}$. The sign of $\Eex$ depends on the FE state.

Reading of data (polarization direction) is achieved by a small reading bias
voltage and a positive gate voltage. The gate voltage should be
smaller than the switching voltage to prevent the polarization switching.
In Sec.~\ref{Sec:Val} we discuss the characteristic time scales.

\subsection{Electric field sensors}

The dependence of conductivity $\sigma$ of granular film on the external electric field can be used in the electric field measuring devices with sensitivity being defined
as follows $\chi=\sigma^{-1}\partial \sigma / \partial \Eex=-2e^2(\epsilon^{\mathrm{\scriptscriptstyle FE}})^{\prime}_{\Eex}/Th(\epsilon+\epsilon^{\mathrm{\scriptscriptstyle FE}})^2$. Below the Curie temperature the field derivative (sensitivity), $\partial \sigma / \partial \Eex$, is finite at zero external field $\Eex$. The sensitivity is defined by the slope of the dependence of dielectric permittivity on the electric field. Therefore
the large sensitivity can be achieved for
the FE substrate with strong field dependent dielectric permittivity.

\subsection{Temperature sensors}

The fact that the FE substrate can enhance the temperature dependence of conductivity of granular film below the FE curie point can be used in temperature measurements. The
temperature sensitivity of such a system can be tuned by an external electric field.

\subsection{Timescales \label{Sec:Val}}

Here we discuss two characteristic timescales in this problem:
1) the electron hopping time between grains, $\tau_{\mathrm{e}} = R_{\mathrm g} C_{\mathrm g}$ with
$R_{\mathrm g}$ and $C_{\mathrm g}$ being the intergrain resistance
and the grain capacitance, respectively, and 2) the polarization switching time, $\taup$.

Our consideration is valid for $\taup\ll\tau_{\mathrm{e}}$ meaning that
the polarization effectively screens the excess electron electric field and the imagine charge has enough
time to form while the excess electron is present at a certain grain. Physically this limit corresponds to
the hopping of coupled pair of electron and the image charge instead of a single electron hopping which
correspond to the opposite limit,  $\taup \gg \tau_{\mathrm{e}}$.
In this case it is necessary to average the interaction of the FE film with the
electric field created by electrons over the large timescale. This leads to the suppression
of screening effect and disappearance of coupling between the FE and the granular film.
For a single electron transistor with the FE this effect was discussed in Ref.~\onlinecite{Beloborodov2014_2}.

For $\taup\ll\tau_{\mathrm{e}}$
the highest possible performance frequency for memory applications is $1/\tau_{\mathrm{e}}$.
\begin{figure}
\includegraphics[width=0.7\columnwidth]{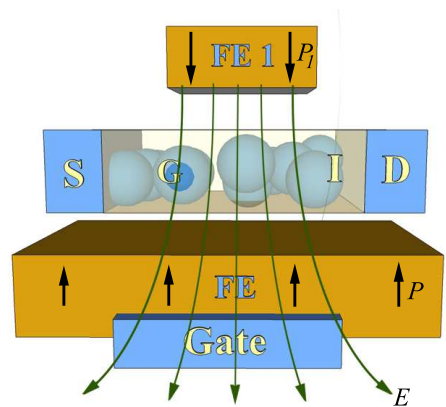}\\
\caption{(Color online) Granular film between two ferroelectrics. Bottom FE stores the
data (polarization direction). Upper FE particle with fixed polarization $P_1$
creates the bias electric field ($E$) for data reading.}\label{Fig_sys2}
\end{figure}

To investigate the system behavior for arbitrary time $\tau_{\mathrm e}$ one
needs to take into account time dispersion of the FE response and solve time dependent
equations for the FE substrate and grains.

Characteristic time $\tau_{\mathrm e}$ varies in a rather large range from dozens of nano- to picoseconds.
This time is controlled by the system geometry and materials.
The FE switching time $\taup$ depends on the material and
can be in the range of $10^{-6}$~s~\cite{Krupanidhi1993PZT} to few nanoseconds.~\cite{Cuppens1991}
Therefore both limits are relevant for experiment.

\subsection{Reading without gate voltage}

For data reading one can use the stray field of the FE particle placed above the
granular film instead of external electric field produced by the gate electrode, Fig.~\ref{Fig_sys2}.
A small FE particle with fixed polarization perpendicular to the film surface creates a stray electric field
acting on the bottom FE layer as an external field, $\Eex$.

\subsection{Ultimate size of granular FE memory}

Reducing the size of granular film one can reach the ultimate size of memory element - a single grain.
In this limit the memory cell is the single electron transistor consisting of a metallic
grain located between two leads. The transport properties of this system were discussed in Ref.~\onlinecite{Beloborodov2014_2}.

\section{Conclusion}

We studied electron transport properties of granular film placed above the FE substrate.
We showed that the conductivity of granular film strongly depends on the FE state due to
screening effects which changes the Coulomb energy in the granular film.
We showed that the FE/granular film system has a large electro-resistance effect.
The conductivity of granular film depends on the direction of the FE polarization.
This effect can be utilized in memory and electric field sensors applications.
We showed that the conductivity of the FE/granular film system strongly depends on temperature in the vicinity
of the FE-paraelectric phase transition. In particular, the conductivity has a peak at the Curie point
in contrast to the usual situation where conductivity decreases in the vicinity of phase transition.
This effect can be used in temperature sensing devices.

\acknowledgments
We thank Michael Huth for providing us with his manuscript prior publication.
N.~C. was supported by RFBR No.~13-02-00579, the Grant of President of Russian Federation for support of Leading Scientific Schools No.~6170.2012.2, RAS presidium and Russian Federal Government programs.
I.~B. was supported by NSF under Cooperative Agreement Award EEC-1160504,
NSF Award DMR-1158666, and NSF PREM Award.

\appendix

\section{Grain charges in equilibrium \label{sec:AppScren}}

Here we discuss the relation between the grain charges $q^{\mathrm g}$ and the applied gate voltage $V_{\mathrm g}$
using the following model: All grains have the same size and the same distance above the FE substrate.
We consider the granular film as an effective medium with dielectric permittivity $\epsilon_{\mathrm{eff}}$
assuming that $\epsilon_{\mathrm{eff}}\approx\epsilon_{\mathrm I}$.

We consider a certain grain $G$ and find the electric field along the shortest line connecting the
grain center and the gate electrode. The field along this line is created by the grain $G$, all other grains,
charges on the FE surface and the gate electrode. We model all grains expect the grain $G$
as a homogeneously charged plane with a hole of radius $r_{\mathrm g}$ around the grain $G$. The
electric field created by the grain $G$ in vacuum is $E_{\mathrm g}=q^{\mathrm g}/z^2$,
where $z$ is the coordinate perpendicular to the FE surface. The origin of the z-axis is in the
center of grain $G$. The electric field created by the charged plane with a hole is give by the superposition
of homogeneous plane and the oppositely charged disk (in vacuum this field is $E_{\mathrm{disk}}=2\pi \sigma^{\mathrm g}(1-z/((z^2+r^2_{\mathrm g})^{1/2}))$), where $\sigma^{\mathrm g}$ is the average surface density of the charges at the grains layer. In the vicinity of the grain $G$ the field is strongly inhomogeneous due
to inhomogeneous charge distribution. Away from the grain, $z\gg r_{\mathrm g}$, the field
turns into the uniform field of charged plane.

The FE substrate is characterized 1) by the polarization $P$ which is perpendicular to
the FE surface and depends on the average electric field inside the FE and 2) by the anisotropic
dielectric permittivity tensor $\hat{\epsilon}^{\mathrm{\scriptscriptstyle FE}}$.
We assume that one of the main axis of the FE is co-directed with the z-axis.
Since the boundary conditions at the FE surface includes only the perpendicular component
of electric induction only the z-component of $\hat{\epsilon}^{\mathrm{\scriptscriptstyle FE}}$ is important.
We denote this component as $\epsFE$. It depends on the electric field applied to the FE.
The uniform polarization of the FE film $P$ leads to the appearance of surface electric
charges on the FE surface $\sigma^{\mathrm P}=\pm P$, where the sign "$\pm$" describes the opposite FE surfaces.

Since the dielectric permittivity of the FE is much larger than the permittivity of the
insulating matrix above the FE substrate the field produced by the grain $G$ and by the opposite charged
disk create the image charges inside the FE film. The electric field inside the insulating layer,
between the grains and the FE substrate, is given by the sum of electric field created by the grains
and their images. The image charges
are $q^{\mathrm g} (\epsilon_{\mathrm I}-\epsFE)/(\epsilon_{\mathrm I}+\epsFE)$. To calculate the
electric field inside the FE we use the effective charge
$q^{\mathrm g} 2\epsilon_{\mathrm I}/(\epsilon_{\mathrm I}+\epsFE)$ for
the grain $G$ charge and the disk charge. We neglect the images appearing due
to the gate since the distance between grains and the gate is large.

We estimate the potential difference between
the grain $G$ and the electrode in the leading approximation as follows
$\Delta \phi\approx 4\pi\sigma^{\mathrm g} (h-a)/\epsilon_{\mathrm I}+4\pi(\sigma^{\mathrm g}+P)L/\epsFE+2\pi\sigma^{\mathrm g}(h-a)\epsFE/(ah\epsilon_{\mathrm I}(\epsilon_{\mathrm I}+\epsFE))$. We use the following set of parameters:
$h-a\approx 1$ nm, $\epsilon_{\mathrm I}\approx 5$, $\epsFE\approx 30$ and the FE
thickness is  $\sim 300$ nm
to reduce the depolarizing electric field inside the FE by the order of magnitude.
For these parameters the screening charges at zero gate voltage are $\sim 0.9 \, P \, r^2_{\mathrm g}$.

The screening of the FE polarization by the grains is suppressed for intergrain distances larger than
the grain size. In addition, the insulating layer between the granular layer and the
FE reduces the screening. The screening is small for FE with high dielectric permittivity, $\epsFE>1000$.

At finite gate voltage the screening charge is given by the expression
\begin{equation}\label{Eq_ScCh}
q^{\mathrm g}=\frac{V_{\mathrm g}-4\pi LP/\epsFE}{\frac{4\pi(h-a)}{\epsilon_{\mathrm I}}+\frac{4\pi L}{\epsFE}+\frac{2\pi(h-a)\epsFE}{(ah\epsilon_{\mathrm I}(\epsilon_{\mathrm I}+\epsFE))}}.
\end{equation}

\section{Coulomb energy\label{Sec:App_Coulomb}}

\subsection{Simplified model}

\begin{figure}
\includegraphics[width=0.5\columnwidth]{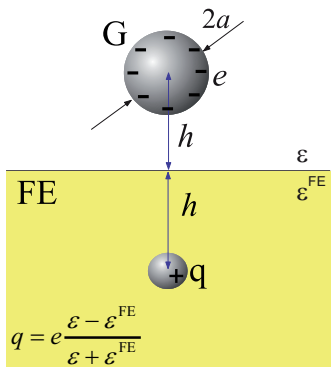}\\
\caption{(Color online) Single metallic grain with size $2a$ and charge $e$ at distance $h$
above the FE substrate. $\epsilon$ is the dielectric permittivity of the medium above the FE substrate.
$\epsilon^{\mathrm{\scriptscriptstyle{FE}}}$ is the dielectric permittivity of the FE substrate. $q$ is the image charge appearing in the FE substrate. Interaction of metallic grain with the image charge modifies the grain charging energy $E_{\mathrm{c}}$.}\label{ImCh}
\end{figure}

Here we discuss the derivation of the charging energy in Eq.~\eqref{Gap},~\cite{Beloborodov2014_1}.
We consider a simple model qualitatively describing the influence of the FE substrate on the Coulomb energy in granular film.
We consider a metallic sphere with radius $a$ and charge $e$ placed above the FE at distance $h$, $a < h$, Fig.~\ref{ImCh}. The external electric field $\Eex$ is applied in the $z$ direction perpendicular to the FE surface. In the
previous section we discussed
the ground state assuming that all grains have the same charge. The ground state
is defined by the mutual capacitance of the whole granular film and the gate electrode. Here we estimate
the electrostatic energy of a single excess electron placed at a certain grain. In this case the capacitance
of a single grain with respect to the whole system (all other grains and the FE substrate) is important.

The FE substrate is characterized by the polarization
vector  $\mathbf{P}$ and the dielectric permittivity tensor $\hat{\epsilon}^{\mathrm{\scriptscriptstyle FE}}$
with $\epsFE$ being its z-component. This description is valid for field created by the grains being
smaller than the switching field of the FE. Both polarization and the permittivity depend on the
external electric field $\Eex$.

Above the FE substrate there is a medium with an effective
dielectric permittivity $\epsilon$ accounting for all other grains and the
insulating matrix. We assume that $\epsilon$ does not depend on the external electric field. The
charged sphere induces the image charge inside the FE substrate.
The magnitude of the image charge is $e (\epsilon-\epsFE)/(\epsilon+\epsFE)$.
Since the dielectric permittivity of the FE is larger than that of the insulator, the image
charge has the opposite sign with respect to the charge of the grain.
The interaction of the charged sphere with the image charge reduces
the electrostatic energy. This effect can be thought of as the effective screening of the
electric field created by electrons on the grains by the FE substrate leading to the reduction
of the charging energy. As a result we obtain the renormalized Coulomb energy, $E_c$ in Eq.~\eqref{Gap}.

\subsection{Coulomb energy in multilayer system}

\begin{figure}
\includegraphics[width=0.6\columnwidth]{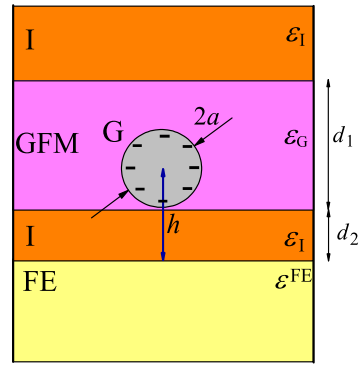}\\
\caption{(Color online) Multilayer system consisting of FE substrate (FE) with
dielectric constant $\epsFE$, insulating layers (I) with dielectric constants $\epsilon_{\textrm{I}}$ and
granular film (GF). Metallic grain (G) of size $2a$ is located above the FE substrate at
distance $h$. Space around the grain in granular film is considered as an
effective medium with dielectric constant $\epsilon_{\textrm{G}}$. Granular film and the middle
insulating layer have thickness $d_1$ and $d_2$, respectively.}\label{Fig:ML}
\end{figure}

In general, the granular system consists of several layers with different
dielectric permittivities. In the simplified model we used the effective dielectric
permittivity $\epsilon$ to describe the layers above the FE substrate. However, this approximation
is valid for the dielectric permittivity of the granular film and the dielectric permittivity of
the insulator below and above the film being close in magnitude.

Below we calculate the charging energy for a grain located in a layered system.
To calculate the charging energy we use the point charges collocation
method.~\cite{Plank2014,Wasshuber2001} The charging energy of metallic grain placed
in a system consisting of several insulating layers (see Fig.~\ref{Fig:ML}) can be estimated as
$E_c=e^2/(2C)$ with $C$ being the grain capacitance.

\begin{figure}
\includegraphics[width=0.9\columnwidth]{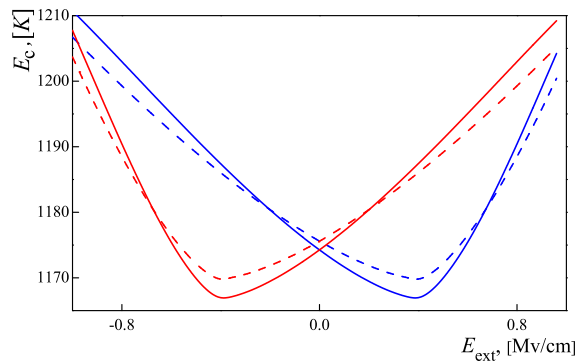}\\
\caption{(Color online) Single grain charging energy placed in layered system shown in Fig.~\ref{Fig:ML}. $d_1=1$ nm, $d_2=7$ nm, $\epsilon_{\mathrm{G}}\approx 1.3$, $\epsilon_{\mathrm{I}}=1$. The
FE is the P(VDF/TrFE) relaxor. Dashed lines correspond to the charging energy calculated using
Eq.~(\ref{Gap}). Solid lines correspond to the numerical method for layered structure.}\label{Fig:CoulLay}
\end{figure}
To find the capacitance $C$ we consider the sphere as the ensemble of point charges $q_i$ placed in the positions $\textrm{r}_i$.
We find charges $q_i$ self-consistently assuming that all the points $\textrm{r}_i$ have
the same potential $\phi_i=\phi$ and the total charge of the sphere is $Q=\sum q_i$. The potential
at points $\textrm{r}_i$ can be found as
\begin{equation}\label{Potential}
\phi_i=\phi=\sum q_j G_{ij},
\end{equation}
where $G_{ij}$ is the electric potential created at point $\textrm{r}_i$ by the unit charge located
at point $\textrm{r}_j$. The Green functions $G_{ij}$ for layered system can be found
using the two dimensional Fourier transformation. The
capacitance $C=Q/\phi$ can be calculated after solving Eqs.~\ref{Potential}.

We calculate the charging energy $E_{\textrm c}$ as a function of electric field
for the following system: the granular film is placed above the P(VDF/TrFE).
The grain radius is $a =3$ nm, the height is $h= 4$ nm ($d_2=1$ nm),
the intergrain distance is $r_{\mathrm g}=7$ nm,
the thickness of the granular film is $d_1 = 7$ nm and
the effective dielectric permittivity is $\epsilon_{\textrm{G}} \approx 1.3$.
The solid lines in Fig.~\ref{Fig:CoulLay} shows the charging energy vs. field $\Eex$ calculated using
numerical method. The dashed lines stand for curves obtained using Eq.~(\ref{Gap}).
The numerical calculations with more complicated model produce almost the same
result as the simplified model. However, the difference increases
with increasing the difference $\epsilon_{\mathrm I}-\epsilon_{\mathrm G}$.

\bibliography{FE_mem}

\end{document}